\documentclass[%
reprint,
 amsmath,amssymb,
 aps,
]{revtex4-2}

\usepackage{color}
\usepackage{graphicx}
\usepackage{dcolumn}
\usepackage{bm}
\usepackage{hyperref}
\usepackage[mathlines]{lineno}
\usepackage{subfigure}

\begin{document}

\title{A model independent approach to the study of $f(R)$ cosmologies with expansion histories close to $\Lambda$CDM}

\author{Saikat Chakraborty}
\email[]{saikatnilch@gmail.com}
\email[]{snilch@yzu.edu.cn}
\affiliation{Center for Gravitation and Cosmology, College of Physical Science and Technology, Yangzhou University, Yangzhou 225009, China
\\
and
\\
International Center for Cosmology, Charusat University, Anand 388421, Gujrat, India}
\author{Kelly MacDevette}
\email[]{mcdkel004@myuct.ac.za}
\affiliation{Department of Mathematics and Applied Mathematics, University of Cape Town, Rondebosch 7700, Cape Town, South Africa}
\author{Peter Dunsby}
\email[]{peter.dunsby@uct.ac.za}
\affiliation{Department of Mathematics and Applied Mathematics, University of Cape Town, Rondebosch 7700, Cape Town, South Africa}
\date{\today}

\begin{abstract}
We propose a new framework for studying the cosmology of $f(R)$ gravity which completely avoids using the reconstruction programme. This allows us to easily obtain a qualitative feel of how much the $\Lambda$CDM 
model differs from other $f(R)$ theories of gravity at the level of linear perturbation theory for theories that share the same background dynamics. This is achieved by using the standard model independent cosmographic parameters to develop a new dynamical system formulation of $f(R)$ gravity which is free from the limitation of having to first specify the functional form of $f(R)$. By considering a set of representative trajectories, which are indistinguishable from $\Lambda$CDM, we use purely qualitative arguments to determine the extent to which these models deviate from the standard model by including an analysis of the linear growth rate of density fluctuations and also whether or not they suffer from the Dolgov-Kawasaki instability. We find that if one demands that a late time $f(R)$ cosmology is observationally close to the $\Lambda$CDM model, there is a higher risk that it suffers from a Dolgov-Kawasaki instability. Conversely, the more one tries to construct a physically viable late time $f(R)$ cosmology, the more likely it is observationally different from the $\Lambda$CDM model. 
\end{abstract}

\maketitle

\section{Introduction}
Assuming that its geometry is well described by a Friedmann-Robertson-Walker metric, there is now considerable observational evidence \cite{Perlmutter:1998np,Riess:1998cb,Tonry:2003zg,Knop:2003iy,Riess:2004nr,Astier:2005qq,Spergel:2003cb,Tegmark:2003ud,Seljak:2004xh,Cole:2005sx,Eisenstein:2005su,Blake:2005jd,Jain:2003tba} supporting the idea that our Universe is currently undergoing an accelerated phase of expansion. Given that gravity is attractive in standard General Relativity (GR) in the presence of both relativistic and non-relativistic matter components, these observations suggest two different possibilities. The first proposes that there exists a matter component which currently dominates the energy density of the Universe, in the presence of which, gravity is repulsive even within standard General Relativity. Because of the lack of a physical description of this matter component, this term in the governing field equations is referred to as  ``Dark Energy".  The simplest such model adds a \emph{cosmological constant} $\Lambda$ to the Einstein-Hilbert Lagrangian, which behaves effectively like a perfect fluid with an equation of state parameter $w=-1$, whose energy density remains constant with time. Together with another unknown matter component called Cold Dark Matter (CDM) this description of the Universe is known as the $\Lambda$CDM model (or \emph{Concordance Model} \cite{Ostriker:1995rn}). $\Lambda$CDM appears to be almost perfectly consistent with current observations. Other 
models in these category introduce either an extra perfect fluid or an extra scalar field to produce a late time accelerating cosmological solution (See Refs. \cite{Copeland:2006wr,Bahamonde:2017ize} for a review on various Dark Energy models). 

The second possibility is that there is no dark energy, but instead, the acceleration of the present Universe is accounted for by a modification of how gravity behaves at very large cosmological scales. This line of thought differs from the first proposal in that it replaces our ignorance of what the dominant matter component is by a lack of understanding of how gravity works at very large distance scales.  Consequently, many so called theories of ``modified gravity" have been developed in order to provide a more geometrical explanation of the late time acceleration of the Universe \cite{Tsujikawa:2010zza,modgrav}. Almost all such models belong to the class of scalar-tensor theories which involve a modification of the Einstein-Hilbert Lagrangian of the following generic form: 
\begin{equation*}
    R \longrightarrow f(R,\phi)\,,
\end{equation*}
while the standard matter part of the Lagrangian remains intact. Careful analysis of the dynamics of such theories show that such theories in general contain two extra scalar degrees of freedom: one is the scalar field $\phi$ and the other is hidden in the form of $\varphi\equiv F(R,\phi)=\frac{\partial f}{\partial R}$. The simplest modified gravity models include only one scalar degree of freedom, which include generalised Brans-Dicke theories ($f(\phi)R$) and $f(R)$ theories. The first ever modified gravity model for dark energy presented in \cite{Carroll:2003wy} was of this type and had the following Lagrangian: 
\begin{equation}
    f(R) = R - \frac{\mu^4}{R}\,,
\end{equation}
although shortly afterwards this particular $f(R)$ theory was shown to be plagued by an instability \cite{Dolgov:2003px}. Later attempts in this line gave birth to several different $f(R)$ gravity models which can describe a late time acceleration of the universe while being free from instabilities, some of the well-known models being the Hu-Sawicki model \cite{Hu:2007nk}, Starobinski model \cite{Starobinsky:2007hu}, MJW model \cite{Miranda:2009rs} etc. In what following we will focus our concern to the $f(R)$ class of modified gravity only.

No matter how many different $f(R)$ gravity models have been proposed as an alternative to $\Lambda$CDM, the $\Lambda$CDM model still remains the one that best fits current observations. The problem with the $\Lambda$CDM model is that the only known candidate for $\Lambda$, namely the quantum vacuum energy as calculated in quantum field theory, has a value many orders of magnitude higher than its observed value. This is the principle motivation for looking for alternative dark energy models. Because $f(R)$ theories present us with an extra dynamical degree of freedom to play with, it is \emph{in principle} possible most of the time to find a class of $f(R)$ theories that can produce the exact same cosmological solution as one finds in General Relativity, \emph{i.e.,} the solution is not unique to GR \cite{Multamaki:2005zs}. The systematic bottom-up approach to finding the class of $f(R)$ gravity models that produces a desired cosmological solution is called a reconstruction programme of $f(R)$ gravity \cite{Nojiri:2009kx}. Even though a number of such reconstruction techniques exist, these methods are not always helpful. More often than not a compact functional form of $f(R)$ cannot be found and even if they can be found,  they involve functions that are too complicated for further analytical treatment. For example, since the $\Lambda$CDM model best fits the observations, there have been attempts to reconstruct $f(R)$ theories that can exactly mimic the $\Lambda$CDM evolution history without invoking a $\Lambda$-term, which gave back compact solutions in the form of Hypergeometric functions \cite{Multamaki:2005zs,Dunsby:2010wg,He:2012rf}. If one tries to analyze further into the perturbative level to look whether perturbation-dependent observables can distinguish between $R-2\Lambda$ and $f(R)$ models, Hypergeometric functions are not the easiest functions to deal with. 

The main purpose of this paper is to completely avoid using the reconstruction programme, while still being able to get a qualitative feel for how much the $\Lambda$CDM model differs at the linear perturbation level from other $f(R)$ theories of gravity, based on late time models that produce the exact same background dynamics. 

A very good approach for qualitative understanding of cosmological models is the dynamical system approach, first developed by Collins \cite{collins} and extensively reviewed in the book edited by Ellis and Wainwright \cite{ellisbook} (see also \cite{dsa_coley,Boehmer:2014vea,Bahamonde:2017ize}). In fact, using these techniques to study cosmological models has the advantage of providing a relatively simple method for obtaining exact solutions, which appear as fixed points of the system, and obtaining a global picture of the dynamics of these models. All the autonomous dynamical system formulations for $f(R)$ gravity that have appeared in literature up to now require one to specify the functional form of $f(R)$ in order to be able to write down a closed system of first order nonlinear ODEs \cite{Amendola:2006we,Carloni:2007br,Guo:2013swa,Carloni:2015jla,Alho:2016gzi}. This approach is therefore not particularly helpful for the general dynamical analysis required in this present work, where the functional form of $f(R)$ is not known \emph{a-priori} (see however \cite{deSouza:2007zpn}). We however circumvent this problem by proposing a novel dynamical system formulation of $f(R)$ gravity that is free from this limitation. The trick is to extend the phase space by including a set of dimensionless cosmographic parameters. Cosmographic parameters are key cosmological observables based on performing a Taylor expansion of the scale factor around the present redshift and, as we will show later on with the specific example of the $\Lambda$CDM case, the choice of a cosmological solution places a constraint on these parameters. It is this constraint which allows one to write down a closed system of first order nonlinear ODEs \emph{without} needing to specify the functional form of $f(R)$. By its very construction this formulation is intimately related to the reconstruction programme and helps us in situations where the reconstruction programme fails to provide a clear picture. To the best of our knowledge this is the first ever proposal for a dynamical systems formulation in $f(R)$ which does not require specifying a functional form of $f(R)$, and therefore links the powerful dynamical system approach to the reconstruction programme.

The paper is organised as follows:
\begin{itemize}
    \item Section \ref{cosm_recon} consists of an overview of $f(R)$ cosmology and the reconstruction programme for $f(R)$ gravity.
    \item Section \ref{DSA} introduces our novel dynamical systems formulation in terms of the cosmographic parameters.
    \item Section \ref{appl_LCDM} applies our dynamical systems formulation to $\Lambda$CDM cosmology as an example.
    \item Section \ref{cosm_pert} gives a brief overview on the calculation of the density contrast parameter at the linear regime of cosmological perturbations.
    \item Section \ref{comparison} details the qualitative comparison between the $\Lambda$CDM model in GR and alternative late time models in $f(R)$ that are cosmographically equivalent at the background level.
    \item We conclude in section \ref{concl} with a summary of the present work and mention some other possible cosmologically relevant research problems that can be addressed along the same line of thought.
\end{itemize}
Throughout the paper we use the $(-+++)$ signature and unit system with $c=1$ and $\kappa=8\pi G=1$.

\section{$f(R)$ cosmology and reconstruction method}
\label{cosm_recon}
In this section we give a very brief overview of cosmology in $f(R)$ gravity. An interested reader is referred to the beautiful topical review on the subject by Sotiriu and Faraoni \cite{Sotiriou:2008rp} or De-Fellice and Tsujikawa \cite{DeFelice:2010aj}. $f(R)$ gravity is characterised by the existence of a propagating scalar degree of freedom $\varphi=F(R)\equiv f'(R)$ as apparent from the trace field equation
\begin{equation}
    RF(R) - 2f(R) + 3\Box F(R) = T\,,
\end{equation}
$T$ being the trace of the energy momentum tensor. General Relativity is the trivial case of $f(R)$ for which $F(R)=1$ and the scalar degree of freedom is no longer propagating. This scalar degree of freedom is sometimes dubbed as the \emph{scalaron}, a term that we will use hereafter. In the presence of a perfect fluid with energy density $\rho$ and pressure $P$, the field equations for $f(R)$ gravity can be expressed as
\begin{subequations}\label{fe}
\begin{eqnarray}
&& 3F\left(H^2 + \frac{k}{a^2}\right) = \rho_{\rm eff} \equiv \rho + \rho_{R}\,,\\
&& -F\left(2\dot{H} + 3H^2 + \frac{k}{a^2}\right) = P_{\rm eff} \equiv P + P_{R}\,,
\end{eqnarray}
\end{subequations}
where $F\equiv\frac{df}{dR}$ and we have defined the scalaron energy density and pressure as
\begin{subequations}
\begin{eqnarray}
&& \rho_{R} \equiv \frac{1}{2}(RF-f) - 3H\dot{F}\,,\\
&& P_{R} \equiv \ddot{F} + 2H\dot{F} - \frac{1}{2}(RF-f)\,.
\end{eqnarray}
\end{subequations}
The effective equation of state parameter of the universe is defined as \begin{equation}\label{w_eff}
    w_{\rm eff} \equiv \frac{P_{\rm eff}}{\rho_{\rm eff}} = \frac{P + P_R}{\rho + \rho_R} = -\frac{2\dot{H} + 3H^2 + k/a^2}{3(H^2 + k/a^2)}\,,
\end{equation}
and the equation of state of the scalaron is
\begin{equation}\label{w_R}
    w_R \equiv \frac{P_R}{\rho_R} = \frac{\ddot{F} + 2H\dot{F} - \frac{1}{2}(RF-f)}{\frac{1}{2}(RF-f) - 3H\dot{F}}\,.
\end{equation}
If the perfect fluid is barotropic, with an equation of state parameter $w$, then $w_{\rm eff}$ and $w_R$ are related to each other via the relation
\begin{equation}\label{www}
    w_{\rm eff} = w\frac{\rho}{\rho_{\rm eff}} + w_R\frac{\rho_R}{\rho_{\rm eff}}\,.
\end{equation}
There are two important conditions for physical viability of any $f(R)$ gravity which we just mention below:
\begin{itemize}
    \item $f'(R)<0$ makes the scalar degree of freedom appearing in the theory a ghost. To eradicate the possibility of a ghost degree of freedom, one must require $f'(R)>0$ for all $R$.
    \item $f''(R)<0$ is related to unstable growth of curvature perturbations in the weak gravity limit (also known as the Dolgov-Kawasaki instability \cite{Dolgov:2003px}). Therefore, one requires that $f''(R)>0$ at least during the early epoch of matter domination.
\end{itemize}
Given a particular $f(R)$ theory of gravity, one can analyse the solutions using the field equations $\eqref{fe}$. However, at times it might be important to find what form of $f(R)$ gravity can give rise to a particular desired solution. This problem is systematically addressed by the \emph{reconstruction} methods of $f(R)$ gravity \cite{Nojiri:2009kx}. There are in fact various reconstruction techniques \cite{Carloni:2010ph}, and it is quite possible that for some particular desired cosmology one of the reconstruction techniques fails while another succeeds to give back a compact form of $f(R)$. The classical and most common reconstruction technique attempts to reconstruct an $f(R)$ from a given solution $a(t)$. The limitation of this technique is that it relies on the invertibility of the function
\begin{equation}
    R(t) = 6(\dot{H}(t) + 2H^2(t))\,.
\end{equation}
If one can find an explicit function $t=g(R)$, then $f(R)$ can be reconstructed by solving the second order differential equation \cite{Carloni:2010ph}
\begin{eqnarray}
3H[g(R)]\dot{R}[g(R)]f''(R) + \left(3H^2[g(R)] - \frac{R}{2}\right)f'(R) && \nonumber\\
+ \frac{1}{2}f(R) = \rho(R) && . 
\end{eqnarray}
Solutions to this equation give a class of $f(R)$ gravity theories for which the given function $a=a(t)$ is an exact solution. If the solution is not given in the compact functional form $a(t)$ or the function $R=R(t)$ is non-invertible, it proves useful to adopt a different reconstruction technique. The whole reconstruction programme relies on expressing the Ricci scalar $R$ as a function of some cosmological variable such that the function is invertible. Alternative reconstruction techniques involve expressing the Ricci scalar as a function of other variables \emph{e.g.} $R=R(a),\,R(H)\,\text{or}\,R(\tau)$ where $\tau\equiv\ln a$. Even if $R(t)$ is non-invertible, $R=R(a),\,R(H)\,\text{or}\,R(\tau)$ \emph{etc.} might be invertible. This is what makes one reconstruction method succeed but the others fail in a particular case.

For example $\Lambda$CDM cosmological evolution is given by the condition \cite{Carloni:2010ph}
\begin{equation}
    \dot{a} = \sqrt{\frac{\rho_0}{a} + \Lambda a^2}\,,
\end{equation}
where $\rho_0$ is a positive constant and $\Lambda>0$ is the cosmological constant. In this case it proves to be much more useful to express the Ricci scalar as either a function of the scale factor $a$ (as in \cite{Dunsby:2010wg}) 
\begin{equation}
    R = \frac{\rho_0}{a^3} + 4\Lambda\,,
\end{equation}
or a function of $\tau\equiv\ln a$ (as in \cite{He:2012rf})
\begin{equation}
    R = \rho_0 e^{-3\tau} + 4\Lambda\,,
\end{equation}
both of which are perfectly invertible. In the first case $f(R)$ is reconstructed from the differential equation \cite{Dunsby:2010wg}
\begin{eqnarray}\label{LCDM_recon_1}
-3(R-3\Lambda)(R-4\Lambda)f''(R) + \left(\frac{R}{2}-3\Lambda\right)f'(R) && \nonumber\\
+\frac{1}{2}f(R) = R - 4\Lambda && \,.
\end{eqnarray}
In the second case $f(R)$ is reconstructed from the following pair of equations
\begin{subequations}\label{LCDM_recon_2}
\begin{eqnarray}
&& \frac{d^2 F[R(\tau)]}{d\tau^2} + \left(\frac{d \ln H}{d\tau} -1\right)\frac{dF[R(\tau)]}{d\tau} \nonumber\\
&& \hspace{20mm} + 2\frac{d \ln H}{d\tau}(F[R(\tau)]-1) = 0\,,\\
&& f(R) = R + \int(F[R(\tau)]-1)R'(\tau)d\tau\,.
\end{eqnarray}
\end{subequations}
 If it is possible to write a reconstruction differential equation, there is no guarantee that a compact form for the general solution can be found. Even in cases where a compact form can be found, it can involve functions that are too complicated for any further analytical treatment (\emph{e.g.,} an analysis of the cosmological perturbations). For example in the $\Lambda$CDM case both the Refs.\cite{Dunsby:2010wg} and \cite{He:2012rf} find the general solution in terms of Hypergeometric functions, whereas the particular solution is the usual $R-2\Lambda$. The whole philosophy of this paper is to avoid solving the reconstruction differential equation, but still be able to qualitatively compare the general solution $f(R)$ against the particular solution $R-2\Lambda$. 

Before moving on to the next section, let us mention here that, as pointed out in Ref.\cite{Chakraborty:2018thg}, even though in most cases a compact form $f(R)$ cannot be found as a solution of the reconstruction differential equation, a series solution in $R$ can always be written (except in rare situations when the point $R=0$ is an irregular singular point of the differential equation). Numerical solution can be obtained provided a pair of boundary conditions. For the $\Lambda$CDM case, setting the boundary conditions $f(R\rightarrow0)=-2\Lambda$, $F(R\rightarrow0)=1$ gives back the unique solution $f(R)=R-2\Lambda$.

\section{A generic dynamical system formulation for $f(R)$ gravity}
\label{DSA}
In terms of expansion normalised dynamical dimensionless variables for $f(R)$ gravity \cite{Carloni:2007br,Bahamonde:2017ize,DeFelice:2010aj},
\begin{eqnarray}\label{dyn_var_def}
& x = \frac{\dot{F}}{HF},\quad y = \frac{R}{6H^2},\quad z = \frac{f}{6FH^2},\\
& \Omega_ = \frac{\rho}{3FH^2},\quad K = \frac{k}{a^2 H^2}\,,   
\end{eqnarray}
the Friedmann constraint equation becomes
\begin{equation}\label{constraint}
 - x + y - z - K + \Omega = 1\,.
\end{equation}
Choosing to eliminate $K$ using the Friedmann constraint, the dynamical system can be expressed as
\begin{subequations}\label{dyn_sys_1}
\begin{eqnarray}
&& \frac{dx}{d\tau} = -4z -2x^2 - (z+2)x +2y +\Omega(x +1 -3w)\,,
\\
&& \frac{dy}{d\tau} = y[2\Omega -2(z-1) + x(\Gamma-2)]\,,
\\
&& \frac{dz}{d\tau} = z(-2z +2\Omega -3x +2) + xy\Gamma\,,
\\
&& \frac{d\Omega}{d\tau} = \Omega(2\Omega -3x -2z -3w -1)\,,
\end{eqnarray}
\end{subequations}
where $\Gamma=\Gamma(R)$ is defined as
\begin{equation}
    \Gamma(R) \equiv \frac{d\ln R}{d\ln F} = \frac{F}{RF'}\,.
\end{equation}
Given a functional form for $f(R)$, one can \emph{in principle} invert the relation
\begin{equation}\label{yz_f}
    \frac{y}{z} = \frac{RF}{f}
\end{equation}
to determine $R=R(y/z)$ and correspondingly find $\Gamma=\Gamma(y/z)$, so as to make the dynamical system \eqref{dyn_sys_1} autonomous. On the other hand, from the definition of $\Gamma$ one can write
\begin{equation}
    xy\Gamma = \frac{\ddot{H}}{H^3} -2(1+y) +6(x+z+\Omega)\,.
\end{equation}
One can keep the term $\ddot{H}/H^3$ as an explicit time dependant term in the system \eqref{dyn_sys_1}, but this makes the system non-autonomous and in general fixed point analysis cannot be applied unless this term either vanishes or is constant. This is the approach taken in \cite{Odintsov:2017tbc}. However the authors in \cite{Odintsov:2017tbc} only considered the very special cases with $\dot{H}=m H^3$ for a constant $m$, for which the dynamical system becomes autonomous. An alternative approach towards the dynamical systems formulation of $f(R)$ gravity can be given by introducing three cosmographic parameters, namely the deceleration, jerk and snap parameters \cite{Dunajski:2008tg}:
\begin{equation}
    q \equiv -\frac{1}{aH^2}\frac{d^2 a}{dt^2}, \qquad j \equiv \frac{1}{aH^3}\frac{d^3 a}{dt^2}, \qquad s \equiv \frac{1}{aH^4}\frac{d^4 a}{dt^2},
\end{equation}
which are related to each other by
\begin{subequations}
\begin{eqnarray}
&& j = 2q^2 + q - \frac{dq}{d\tau}\,,
\\
&&  s = \frac{dj}{d\tau} - j(2 + 3q)\,.
\end{eqnarray}
\end{subequations}
Since $f(R)$ gravity is a fourth order theory of gravity, i.e., the field equations contain terms including up to fourth derivatives of the metric, higher order cosmographic parameters cannot be included in the field equations. Now one can replace $\Gamma$ in terms of the cosmographic parameters using
\begin{equation}\label{xy_Gamma}
    xy\Gamma = -2(1+y) +6(x+z+\Omega) + j + 3q - 2\,.
\end{equation}
The definition of the Ricci scalar provides an additional constraint equation
\begin{eqnarray}\label{ricci_const}
y = \frac{\dot{H}}{H^2} +2 +K = 1 -q +K\,.
\end{eqnarray}
The Friedmann constraint \eqref{constraint} can therefore be written as
\begin{equation}\label{fried_const}
    z = -x +\Omega -q\,.
\end{equation}
The two constraint equations \eqref{ricci_const} and \eqref{fried_const} can be used to eliminate $y$ and $z$ and write the dynamical system in terms of $x,\,\Omega,\,K,\,q,\,j$:
\begin{subequations}\label{dyn_sys_2}
\begin{eqnarray}
&& \frac{dx}{d\tau} = -x(x -q) +2(x +K +q) -3\Omega(1 +w) +2\,,
\nonumber\\
&& 
\\
&& \frac{d\Omega}{d\tau} = -\Omega(x -2q +1 +3w)\,,
\\
&& \frac{dK}{d\tau} = 2qK\,,
\\
&& \frac{dq}{d\tau} = 2q^2 +q -j\,,
\\
&& \frac{dj}{d\tau} = j(2 +3q) +s\,.
\end{eqnarray}
\end{subequations}
The system of equations \eqref{dyn_sys_2} has a much simpler form as compared to the system of equations \eqref{dyn_sys_1} and also does not require us to explicitly specify the functional form of the underlying $f(R)$ gravity to make the system autonomous. The last two equations are completely kinematical in nature as they depend only on \emph{how} the universe evolves and not at all on \emph{what} inherent dynamics cause the universe to evolve in that way. In general, a particular type of time evolution of the universe can be specified in terms of cosmographic parameters, which allows us to either find the underlying $f(R)$ gravity by solving the reconstruction differential equation \cite{Carloni:2010ph,Capozziello:2008qc}, or to write a closed autonomous dynamical system without finding the explicit solutions. Before proceeding further some comments are in order:
\begin{itemize}
    \item The present universe is accelerating, \emph{i.e.,} the deceleration parameter $q<0$. On the other hand one always has $|K|\leq 1$, \emph{i.e.} $(1+K)\geq 0$. Therefore from the purely kinematic relation \eqref{ricci_const} we can conclude that $y>0$ in our present universe. $y=0$ is an invariant submanifold of the dynamical system \eqref{dyn_sys_1}, \emph{i.e.,} it divides the whole phase space into two disjoint regions with $y>0$ ($R>0$) and $y<0$ ($R<0$) respectively. All the physically relevant dynamics that lead to the present day accelerating universe must take place within the region $y>0$. Existence of an invariant submanifold also implies there cannot be a global attractor or repeller, unless it lies on the invariant submanifold itself.
    \item As mentioned in the previous section, the physical viability of any $f(R)$ gravity requires $F>0$ throughout the physically relevant region of the phase space and $F'\geq 0$ ($F'=0$ corresponding to the special case $f(R)=R+\Lambda$) at least in the neighbourhood of the fixed point corresponding to the matter dominated epoch. Eliminating $y$ and $z$ from \eqref{xy_Gamma} using the constraints \eqref{ricci_const}, \eqref{fried_const} and then using the definition of $\Gamma$ one can write
    \begin{equation}
        \frac{1}{y\Gamma} = \frac{6F'H^2}{F} = \frac{x}{12\Omega -2K -q +j -6}\,.
    \end{equation}
    Assuming the condition $F>0$ is met, demanding $F'\geq 0$ puts the following constraint on the phase phase:
    \begin{equation}\label{phys_const}
        \frac{x}{12\Omega -2K -q +j -6} \geq 0\,.
    \end{equation}
    The submanifold $x=0$ correspond to the GR limit ($F'=0$). 
\end{itemize}

For any fixed point $P$ given by the coordinates
\begin{equation*}
    P \equiv (x^*,y^*,z^*,\Omega^*,K^*,q^*,j^*)\,,
\end{equation*}
the effective equation of state parameter \eqref{w_eff} can be expressed in terms of the dynamical quantities as
\begin{equation}\label{w_eff_dyn}
    w_{\rm eff} = \frac{2q^*}{3(1+K^*)}-\frac{1}{3}\,\qquad (K^*\neq -1)\,.
\end{equation}
The relation between different equation of state parameters (\eqref{www}) can also be expressed in terms of the dynamical quantities
\begin{equation}\label{www_dyn}
    w_{\rm eff} = w\frac{\Omega}{1+K^*} + w_R\frac{y^* -z^* -x^*}{1+K^*} = w_R + (w - w_R)\frac{\Omega^*}{1+K^*}\,,
\end{equation}
where in the last step we have used the Friedmann constraint \eqref{fried_const}. The case $K=-1$ corresponds to a Milne solution. In the vicinity of the fixed point one can \emph{approximately} reconstruct a form of $f(R)$ from the relation
\begin{equation}
    \dot{H} = h(H) \equiv -(1+q^*)H^2
\end{equation}
using the method detailed in Ref.\cite{Carloni:2010ph}. It should however be kept in mind that the $f(R)$ form so obtained is valid only during the epoch represented by the fixed point $P$, and not throughout the whole evolution history that we are considering. 
\section{Application to $\Lambda$CDM cosmology}
\label{appl_LCDM}
As a simple example let us take the observationally successful $\Lambda$CDM cosmological model, where acceleration of the universe is due to a positive cosmological constant $\Lambda$ ($w=-1$) and cold dark matter (CDM) is modelled by a dust-like fluid ($w=0$). The equations of motion are
\begin{subequations}
\begin{eqnarray}
&& 3\left(H^2 + \frac{k}{a^2}\right) = \rho + \Lambda\,,\\
&& 2\dot{H} + 3H^2 + \frac{k}{a^2} = \Lambda\,,\\
&& \dot{\rho} + 3H\rho = 0\,.
\end{eqnarray}
\end{subequations}
As the first and simplest model for late time cosmology, $\Lambda$CDM dynamics are very well studied (\cite{Bahamonde:2017ize,Copeland:2006wr,GoliathEllis}). In terms of the expansion normalised dimensionless dynamical quantities
\begin{equation}
    \Omega_m=\frac{\rho}{3H^2}\,, \qquad \Omega_{\Lambda}=\frac{\Lambda}{3H^2}\,, \qquad K=\frac{k}{a^2 H^2}\,,
\end{equation}
so that the constraint equation becomes 
\begin{equation}
    \Omega_m - K = 1 - \Omega_{\Lambda}\,.
\end{equation}
Using the constraint equation, we choose to eliminate $\Omega_{\Lambda}$. The dynamical system for the $\Lambda$CDM model can then be written as
\begin{subequations}\label{dyn_sys_LCDM}
\begin{eqnarray}
&& \frac{d\Omega_m}{d\tau} = -\Omega_m (2K- 3\Omega_m +3)\,,\\
&& \frac{dK}{d\tau} = -K(2K- 3\Omega_m +2)\,.
\end{eqnarray}
\end{subequations}
For any fixed point $Q$ given by the coordinates
\begin{equation*}
    Q \equiv (\Omega_{\Lambda}^*,\Omega_m^*,K^*)\,,
\end{equation*}
the effective equation of state parameter is
\begin{equation}
    w_{\rm eff} = -\frac{\Omega_{\Lambda}^*}{1+K^*} = \frac{\Omega_m^*}{1+K^*} - 1 \,\qquad (K\neq -1)\,.
\end{equation}
Fixed points of the above system, along with their nature under a linear stability analysis, are listed in Table \ref{tab:1}.
\begin{table}[!h]
    \centering
    \begin{tabular}{|c|c|c|c|}
        \hline 
        Fixed & Coordinates & Stability & Cosmological solution \\
        Point & $(\Omega_m^*,K^*)$ & Nature & \\
        \hline
        $Q_1$ & $(0,0)$ & Attractor & $\Lambda$-dominated De-Sitter \\
        &  &  & ($w_{\rm eff}=w_{\Lambda}=-1,\,H=const.$) \\
        \hline
        $Q_2$ & $(1,0)$ & Repeller & Matter dominated power law \\
        &  &  & ($w_{\rm eff}=w_m=0,\,a~t^{2/3}$) \\
        \hline
        $Q_3$ & $(0,-1)$ & Saddle & Milne solution ($a~t$) \\
        \hline
    \end{tabular}
    \caption{Fixed points of the dynamical system \ref{dyn_sys_LCDM}.}
    \label{tab:1}
\end{table}
The whole $\Lambda$CDM cosmology can be specified by the cosmographic requirement \cite{Dunajski:2008tg,Capozziello:2008qc}
\begin{equation}\label{LCDM}
   j = K +1\,.
\end{equation}
Instead of attempting to reconstruct an $f(R)$ theory that can exactly mimic the $\Lambda$CDM evolution history at the observational level, we rather investigate some of the generic dynamical features of such $f(R)$ theories, and look for any possible deviations from the $\Lambda$CDM cosmological model. Using the observational requirement \eqref{LCDM} the dynamical system for any generic value of the spatial curvature parameter $k$ becomes
\begin{subequations}\label{dyn_sys_3}
\begin{eqnarray}
&& \frac{dx}{d\tau} = -x(x -q) +2(x +K +q) -3\Omega +2\,,
\\
&& \frac{d\Omega}{d\tau} = -\Omega(x -2q +1)\,,
\\
&& \frac{dK}{d\tau} = 2qK\,,
\\
&& \frac{dq}{d\tau} = 2q^2 +q -K -1\,.
\end{eqnarray}
\end{subequations}
Since CDM is modelled by a dust fluid ($w=0$), Eq.\eqref{www_dyn} simplifies to
\begin{equation}
    w_{\rm eff} = w_R \left(1-\frac{\Omega^*}{1+K^*}\right)
\end{equation}
where a `*' denotes the value of the corresponding quantity at a fixed point. Fixed points of the above system, along with their nature under a linear stability analysis, are listed in Table \ref{tab:2}.
\begin{table}[!h]
    \centering
    \begin{tabular}{|c|c|c|c|}
        \hline 
        Fixed & Coordinates & Stability & Cosmological solution \\
        Point & $(x^*,\Omega^*,K^*,q^*)$ & Nature & \\
        \hline
        $P_1$ & $(0,0,0,-1)$ & Saddle & scalaron dominated De-Sitter \\
        &  &  & ($w_{\rm eff}=w_R=-1,\,H=const.$) \\
        \hline
        $P_2$ & $(1,0,0,-1)$ & Attractor & scalaron dominated De-Sitter \\
        &  &  & ($w_{\rm eff}=w_R=-1,\,H=const.$) \\
        \hline
        $P_3$ & $(0,1,0,\frac{1}{2})$ & Saddle & Matter dominated power law \\
        &  &  & ($w_{\rm eff}=w=0,\,a~t^{2/3}$) \\
        \hline
        $P_4$ & $(\frac{5-\sqrt{73}}{4},0,0,\frac{1}{2})$ & Repeller & scalaron dominated power law \\
        &  &  & ($w_{\rm eff}=w_R=0,\,a~t^{2/3}$) \\
        \hline
        $P_5$ & $(\frac{5+\sqrt{73}}{4},0,0,\frac{1}{2})$ & Saddle & scalaron dominated power law \\
        &  &  & ($w_{\rm eff}=w_R=0,\,a~t^{2/3}$) \\
        \hline
        $P_6$ & $(0,0,-1,0)$ & Saddle & Milne solution ($a~t$) \\
        \hline
        $P_7$ & $(2,0,-1,0)$ & Saddle & Milne solution ($a~t$) \\
        \hline
    \end{tabular}
    \caption{Fixed points of the dynamical system \eqref{dyn_sys_3}.}
    \label{tab:2}
\end{table}
$K=0$ is an invariant submanifold of the system \eqref{dyn_sys_3} on which the spatially flat dynamics take place. Fixed points that reside on the $K=0$ submanifold are $P_1$, $P_2$, $P_3$, $P_4$ and $P_5$. On this invariant submanifold one can reduce the phase space even more
\begin{subequations}\label{dyn_sys_4}
\begin{eqnarray}
&& \frac{dx}{d\tau} = -x(x -q) +2(x +q) -3\Omega +2\,,
\\
&& \frac{d\Omega}{d\tau} = -\Omega(x -2q +1)\,,
\\
&& \frac{dq}{d\tau} = 2q^2 +q -1\,.
\end{eqnarray}
\end{subequations}
The system \eqref{dyn_sys_4} represents an autonomous dynamical system corresponding to all possible solutions of the reconstruction equation \eqref{LCDM_recon_1} or \eqref{LCDM_recon_2}. This allows us to qualitatively compare the $R-2\Lambda$ gravity against other possible $f(R)$ theories that produce the same dynamics. We notice that for the spatially flat case the $q$-equation decouples which leads to two new invariant submanifolds: a submanifold $q=-1$ consisting of accelerated cosmological solutions and a submanifold $q=\frac{1}{2}$ consisting of decelerated cosmological solutions. The fixed points $P_1$ and $P_2$ reside on the $q=-1$ submanifold and the fixed points $P_3$, $P_4$ and $P_5$ reside on the $q=\frac{1}{2}$ submanifold. Linear stability analysis reveals that the ``deceleration submanifold'' $q=\frac{1}{2}$ is a repelling one while the ``acceleration submanifold'' $q=-1$ is an attracting one, which is consistent with the fact that $P_2$ is an attractor while $P_4$ is a repeller. 

Another important thing to notice is that unlike the $\Lambda$CDM model where the matter dominated fixed point is a past attractor, in this case the matter dominated fixed point is only a saddle, \emph{i.e.} an intermediate epoch. The true past attractor in this case is the fixed point $P_4$ which represents a scalaron dominated epoch but, surprisingly, the scalaron itself behaves like a dust as far as cosmological dynamics is concerned.

Given the coordinates of a fixed point, it is a straightforward exercise to check whether it satisfies the condition \eqref{phys_const}. In Table \ref{tab:2} the fixed points $P_1$, $P_3$ and $P_6$ lie on the submanifold $x=0$ that represents the GR limit. Therefore these three points satisfy the condition \eqref{phys_const} trivially. It can be easily checked that out of the other four fixed points only the past attractor $P_4$ satisfies the constraint. This fixed point corresponds to a scalaron dominated cosmology but, surprisingly, the scalaron itself behaves like a dust fluid ($w_{\rm eff}=w_R=0$). Therefore this fixed point produces a time evolution exactly the same as that of a matter dominated epoch in GR ($a\sim t^{2/3}$). In particular it should be noted that the scalaron dominated De-Sitter future attractor $P_2$ does not satisfy the condition \eqref{phys_const}. This allows us to conclude that even if there exist possible $f(R)$ models which are able to give rise to cosmological dynamics that are observationally indistinguishable from $\Lambda$CDM dynamics (at least at the background level), the $f(R)$-dynamics will inevitably lead to an epoch where the condition \eqref{phys_const} is not met.

Before leaving this section it is worth mentioning that there are indeed perfectly viable $f(R)$ gravity models that can successfully reproduce transition from a matter dominated power law evolution epoch to a scalaron dominated De-Sitter epoch, while always satisfying the condition \eqref{phys_const}. Two such examples are \emph{e.g.} $e^{\Lambda R}$($\Lambda>0$) \cite{Abdelwahab:2007jp} and $R+\alpha R^n$($\alpha,n>0$) \cite{Abdelwahab:2011dk}. The point we make here is that none of them can mimic the exact evolution history as produced by the $\Lambda$CDM model, which till now remains the most observationally fitted model for late time cosmology. On the other hand if one demands an $f(R)$ model of late time cosmology to reproduce the same expansion history as that of the $\Lambda$CDM model, one is bound to end up in a region where the condition \eqref{phys_const} is no longer satisfied.
\section{Behaviour of cosmological perturbations}
\label{cosm_pert}
It has been suggested \cite{DeFelice:2010aj,Lee:2017lud} that one can distinguish between the $\Lambda$CDM model and an equivalent scalar model by taking into account observables that depend on cosmological perturbations. 
In this section we briefly review this point. For the sake of simplicity we constrain ourselves to the spatially flat case in this section. Given the fact that present day observation suggests that our universe is very close to being spatially flat, it is worthwhile considering this case a little deeper. The perturbation quantity that is of observational interest in late time cosmology is the matter density contrast $\delta\equiv\frac{\delta\rho}{\rho}$. The evolution of this quantity at the sub-Hubble limit is approximately governed by the second order differential equation \cite{DeFelice:2010aj,Lee:2017lud}
\begin{equation}
\frac{d^2 \delta}{d\tau^2} + \left(2+\frac{\dot{H}}{H^2}\right)\frac{d\delta}{d\tau} = \frac{2}{3}\frac{\rho\delta}{FH^2}\left(\frac{1 + \frac{a^2}{4k^2}\frac{F}{F'}}{1 + \frac{a^2}{3k^2}\frac{F}{F'}}\right)\,,    
\end{equation}
with $k$ here being the wavenumber of a particular Fourier mode of $\delta$, not to be confused with the spatial curvature parameter. In terms of the dynamical variables \eqref{dyn_var_def} and utilising the relation \eqref{xy_Gamma}, the term on the right hand side of the above equation can be expressed as
\begin{eqnarray}
\frac{2}{3}\frac{\rho\delta}{FH^2}\left(\frac{1 + \frac{a^2}{4k^2}\frac{F}{F'}}{1 + \frac{a^2}{3k^2}\frac{F}{F'}}\right) &=& 2\Omega\left(\frac{x + \frac{3}{2}xy\Gamma\left(\frac{aH}{k}\right)^2}{x + 2xy\Gamma\left(\frac{aH}{k}\right)^2}\right)\delta \nonumber
\\
&=& \Omega\left(\frac{2x + 3(12\Omega-q-5)\left(\frac{aH}{k}\right)^2}{x + 2(12\Omega-q-5)\left(\frac{aH}{k}\right)^2}\right)\delta \nonumber \,.
\end{eqnarray}
Therefore the sub-horizon perturbation equation can be written as
\begin{equation}\label{ptbn_eqn}
\frac{d^2 \delta}{d\tau^2} + (1-q)\frac{d\delta}{d\tau} - \Omega\left(\frac{2x + 3(12\Omega-q-5)\left(\frac{aH}{k}\right)^2}{x + 2(12\Omega-q-5)\left(\frac{aH}{k}\right)^2}\right)\delta = 0\,.
\end{equation}
Keeping in mind that $k \gg aH$ in sub-horizon limit, Eq.\eqref{ptbn_eqn} can be solved at two different regimes of interest:
\begin{itemize}
    \item \emph{GR regime}: $0\lesssim|x|\ll\left(\frac{aH}{k}\right)^2$: This corresponds to the limit when the modification of gravity theory can be safely ignored. In this limit Eq.\eqref{ptbn_eqn} reduces to
    \begin{equation}\label{ptbn_eqn_1}
        \frac{d^2 \delta}{d\tau^2} + (1-q)\frac{d\delta}{d\tau} - \frac{3}{2}\Omega\delta = 0\,.
    \end{equation}
    \item \emph{$f(R)$ regime}: $0\lesssim\left(\frac{aH}{k}\right)^2\ll|x|$: This corresponds to a limit when the effect of the gravity modification cannot be ignored. In this limit Eq.\eqref{ptbn_eqn} reduces to
    \begin{equation}\label{ptbn_eqn_2}
        \frac{d^2 \delta}{d\tau^2} + (1-q)\frac{d\delta}{d\tau} - 2\Omega\delta = 0\,.
    \end{equation}
\end{itemize}
A matter dominated power law evolution epoch is characterised by $q=\frac{1}{2}$ and $\Omega_m=1$. For the $\Lambda$CDM model only the GR regime is relevant. Solving Eq.\eqref{ptbn_eqn_1} in this regime we get two modes:
\begin{eqnarray}
    & \delta \sim e^{\tau} \sim a \qquad (\text{Growing mode}) \\
    & \delta \sim e^{-\frac{3}{2}\tau} \sim a^{-\frac{3}{2}} \qquad (\text{Decaying mode})
\end{eqnarray}
For $f(R)$ models however both these regimes are possible. In particular, depending on the $f(R)$ theory, it is possible for wavenumbers $k$ relevant to large scale structure observations to transit from the GR regime into the $f(R)$ regime within the matter dominated regime. Solving Eq.\eqref{ptbn_eqn_1} in regime $II$, we get two modes:
\begin{eqnarray}
    & \delta \sim e^{\frac{1}{4}(\sqrt{33}-1)\tau} \sim a^{\frac{1}{4}(\sqrt{33}-1)} \quad (\text{Growing mode})\,,\\
    & \delta \sim e^{-\frac{1}{4}(\sqrt{33}+1)\tau} \sim a^{-\frac{1}{4}(\sqrt{33}+1)} \quad (\text{Decaying mode})\,.
\end{eqnarray}
The growing mode solution is related to the growth rate of large scale structures in the universe, which is a cosmological observable. Clearly there is a difference between the growing mode solutions in the GR regime and $f(R)$ regime. In particular the $f(R)$ regime leads to a faster growth of structures. It is precisely the existence of this regime that leads to an observable difference between the $\Lambda$CDM model and late time $f(R)$ models. 

For a matter perturbation mode of wavelength $\lambda\sim 1/k$, transition from the GR regime to the $f(R)$ regime occurs at a time which is approximately given by
\begin{equation}\label{transition}
    |x| \simeq \left(\frac{aH}{k}\right)^2 ~ \left(\frac{\lambda}{R_c}\right)^2\,,
\end{equation}
where $R_c=\frac{1}{aH}$ is the comoving Hubble horizon. The behaviour of linear matter perturbations therefore depends on the background cosmological evolution. From the definition of $x$ \eqref{dyn_var_def} it is clear that this transition scale depends on the form of $f(R)$ as well. The observable difference in the growth rate of structures between $\Lambda$CDM and $f(R)$ based models depends on this transition scale for a typical perturbation mode $k$ relevant to the large scale structures of the Universe. In the phase space picture, one can argue that if this transition occurs very far from the matter dominated fixed point, then the observable difference from the $\Lambda$CDM model will be negligible. On the other hand if the transition occurs very near to the matter dominated fixed point, then the observable difference from the $\Lambda$CDM model may be significant.
\section{Comparison between $\Lambda$CDM model and equivalent late time $f(R)$ models}
\label{comparison}
\begin{figure}
    \subfigure[\label{fig:1a}]{\includegraphics[scale=0.38]{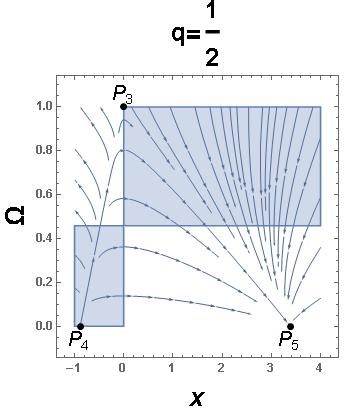}}
    \subfigure[\label{fig:1b}]{\includegraphics[scale=0.38]{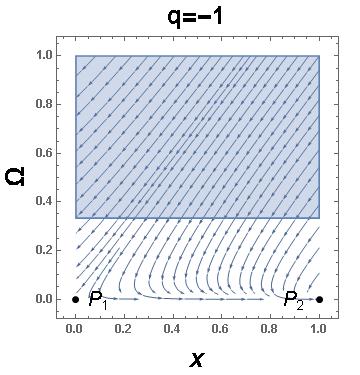}}
    \subfigure[\label{fig:1c}]{\includegraphics[scale=0.38]{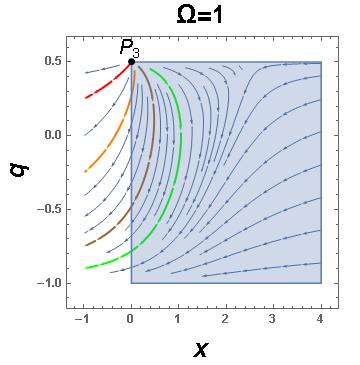}}
    \subfigure[\label{fig:1d}]{\includegraphics[scale=0.38]{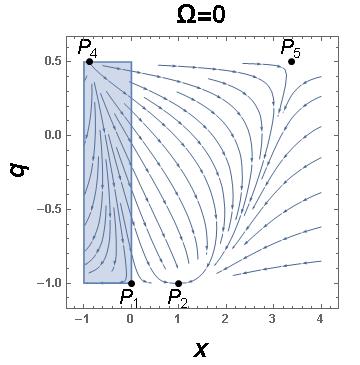}}
    \caption{Projection of the phase portrait on (a) the invariant submanifold $q=\frac{1}{2}$, (b) the invariant submanifold $q=-1$, (c) the slice $\Omega=1$, (d) the slice $\Omega=0$. The shaded region represents the region in which the condition $F'>0$ is satisfied (provided $F$ is already positive).}
    \label{fig:1}
\end{figure}
The phase space corresponding to the dynamical system \eqref{dyn_sys_4} is 3-dimensional. One can get a fairly good idea of the phase space dynamics by considering ``projections" of the phase portrait on different planes. Fig.\ref{fig:1} shows four sets of such projections, along with the labelled fixed points. Figs.\ref{fig:1a} and \ref{fig:1b} show the projections of the phase portrait on the ``deceleration submanifold" $q=\frac{1}{2}$ and the acceleration submanifold $q=-1$ respectively, whereas Figs.\ref{fig:1c} and \ref{fig:1d} show the projections of the phase portrait on the slices $\Omega=1$ and $\Omega=0$ respectively. In the figures, the line $x=0$ represents the GR limit and the region in which the condition \eqref{phys_const} is satisfied is shaded. As one clearly sees, both the fixed points that correspond to accelerating epochs lie outside the shaded region. The most important figure in the context of assessing the observational difference between $\Lambda$CDM and $f(R)$ models is Fig.\ref{fig:1c}, which shows trajectories emanating from the matter dominated fixed point $P_3$ and moving towards the acceleration submanifold $q=-1$. These trajectories represent possible evolution routes for the transition from a matter dominated decelerating epoch to a late time accelerating epoch.   

Let us analyse Fig.\ref{fig:1c} in a little more detail. 
It is clear from the figure that in the vicinity of the matter dominated fixed point $P_3$, the evolution of the phase trajectories is such that $|x|$ increases with time. On the other hand, it is straightforward to calculate that
\begin{equation}
    \frac{d(1/R_c^2)}{d\tau} = -\frac{2q}{R_c^2}\,,
\end{equation}
so that $(1/R_c^2)$ is a decreasing function of time near the fixed point $P_3$. This means that even if a relevant perturbation mode of wavelength $\lambda$ is a-priori within the GR regime ($|x|\ll(\lambda/R_c^2)$), it is possible during the course of cosmic evolution to achieve the transition scale given by Eq.\eqref{transition}, beyond which it enters the $f(R)$ regime ($|x|\gg(\lambda/R_c^2)$). Moreover, one can derive from a straightforward calculation that
\begin{equation}\label{transition1}
    \frac{d}{d\tau}\ln\left(|x|R_c^2\right) = 2q + \frac{(1+q)(1-2q)}{|x|}\frac{d|x|}{d(-q)}\,.
\end{equation}
Within the region between the submanifolds $q=\frac{1}{2}$ and $q=-1$, $q$ is monotonically decreasing and the quantity $(1+q)(1-2q)$ is positive throughout. Moreover, near the fixed point $P_3$ both $q$ and $\frac{d|x|}{d(-q)}$ are positive. Therefore from Eq.\eqref{transition1} one can conclude that the quantity $|x|R_c^2$ is increasing with time near the point $P_3$, \emph{i.e.,} the perturbation modes are moving from the GR regime towards the $f(R)$ regime. It is also clearly seen from Eq.\eqref{transition1} that the rate of change of the quantity $|x|R_c^2$ along a phase trajectory nearby $P_3$ directly depends on the slope of the phase trajectory $\frac{d|x|}{d(-q)}$. The more is the slope $\frac{d|x|}{d(-q)}$, the faster is the increment of the quantity $|x|R_c^2$ and consequently quicker is the transition from GR to $f(R)$ regimes for the perturbation modes. In the paragraph below we discuss this point by taking as example four characteristic phase trajectories.

In the figure have highlighted four characteristic phase trajectories emanating from $P_3$ in colours red, orange, brown and green respectively. These trajectories represent segments of four possible cosmic evolutions, each of which goes though a matter dominated decelerating epoch into an accelerating phase, and is observationally indistinguishable from the $\Lambda$CDM model at the background level. How much they will observationally deviate from the $\Lambda$CDM model at the perturbative level and how physically viable they will be can be qualitatively assessed by carefully examining the phase portrait. From our discussion in the the last paragraph we know that the faster a phase trajectory moves away from the $x=0$ line (\emph{i.e.} has a steeper slope $\frac{d|x|}{d(-q)}$), the quicker is the transition from GR regime to $f(R)$ regime for some particular characteristic perturbation mode of wavelength $\lambda\sim 1/k$, and moreover the corresponding cosmic evolution is expected to observationally deviate from $\Lambda$CDM. On the other hand since the shaded region represents the region where the condition \eqref{phys_const} is satisfied, the faster a phase trajectory goes out of this region, the more the corresponding cosmic evolution is expected to encounter the Dolgov-Kawasaki instability, hence being more physically non-viable.

We note the following:
\begin{itemize}
    \item The leftmost highlighted trajectory (in red) moves away from both the $x=0$ line much faster compared to the other three highlighted trajectories, and does not stay within the shaded region at all in the vicinity of the matter dominated fixed point. Therefore this and nearby trajectories represent a class of cosmic evolutions that is expected to show significant observational deviation from the $\Lambda$CDM, while also being severely plagued by the Dolgov-Kawasaki instability. We can therefore rule out such cosmic evolutions from being physically viable.
    \item The highlighted trajectory second from the left (in orange) stays near the $x=0$ line a little longer compared to the leftmost red trajectory, but spends only a short time within the shaded region. This and nearby trajectories represent a class of cosmic evolutions which are observationally closer to the $\Lambda$CDM model (compared to the red trajectory), but we still cannot characterise them as physically viable enough.
    \item The highlighted trajectory second from the right (in brown) always stays closer to the $x=0$ line, while also being within the shaded region compared to most other trajectories. This and nearby trajectories represent the most optimum cosmic evolutions one can get because of their observable closeness with $\Lambda$CDM and that they avoid the Dolgov-Kawasaki instability.
     \item Finally, the rightmost highlighted trajectory (in green) stays within the shaded region longer than all the other three trajectories, but also moves away from the $x=0$ line almost as fast as the red trajectory. Therefore this trajectory and those near it represent a class of cosmic evolutions that can be safely assumed to be free from the Dolgov-Kawasaki instability, but are expected to show significant observational deviation from $\Lambda$CDM.
\end{itemize}
Figs. \ref{fig:1a} and \ref{fig:1d} show that all trajectories that pass near the saddle point $P_3$ must end up at the future attractor $P_5$. Since $P_5$ lies outside the shaded region, as is clear from Figs.\ref{fig:1b} and \ref{fig:1d}, all such trajectories necessarily end up in a region where the condition \eqref{phys_const} is violated. The trajectories emanating from a region close to $P_3$ that do not remain in the shaded region at all (\emph{e.g.} the red one in Fig.\ref{fig:1c}) can be immediatly discarded as being physically non-viable. Among the other trajectories in Fig.\ref{fig:1c} we notice two competing tendencies. The more a trajectory wants to spend time within the shaded region, the faster it has to deviate from the $x=0$ line (\emph{e.g.} compare the brown and the green trajectory). This result, albeit a qualitative one, is nonetheless quite interesting. Physically this means that the more one demands that a late time $f(R)$ cosmology should be observationally close to the $\Lambda$CDM model, the higher the risk is that it is physically non-viable. On the other hand, the more one tries to construct a physically viable late time $f(R)$ cosmology, the higher the risk of it being observationally different from the $\Lambda$CDM model. This statement is very generic as it is independent of the functional form of $f(R)$; the only constraint being it should be observationally indistinguishable from $\Lambda$CDM at the background level. That such a very generic statement can be made from a purely qualitative phase space analysis is truly remarkable.
\section{Conclusion}
\label{concl}
In this paper we developed a new dynamical systems framework for studying the cosmology of $f(R)$ gravity which completely circumvents the reconstruction programme. This is achieved by using cosmographic parameters to write $f(R)$ cosmology in such a way that it is theory independent. The use of cosmographic parameters as dynamical variables gives rise to a set of algebraic constraints on the phase space which are fixed by observations. All earlier autonomous dynamical system formulations of $f(R)$ gravity require one to specify the form of $f(R)$ to close the system of dynamical equations.
To the best of our knowledge, this is the first time an autonomous dynamical system formulation of $f(R)$ gravity is presented that is \emph{model independent}. 

By considering the qualitative properties of the resulting phase space and the growth rate of matter perturbations, we found that models that are observationally close to $\Lambda$CDM suffer from a higher risk that they encounter a Dolgov-Kawasaki instability in their future. On the other hand,  demanding that such instabilities should not occur leads to trajectories which are very different from the standard model. 

Other well known examples, such as the Hu and Sawicki $f(R)$ theories of gravity \cite{Hu:2007nk} can also be studied using this approach. As far as we are aware only a special case of this theory has been considered using a dynamical systems approach \cite{Sulona}, for which it is possible to write the function $\Gamma$ in terms of the dynamical systems variables. This method should allow for a much more general analysis of the background dynamics of such models and their parameter space. In Ref.\cite{Sulona} it was found for example that great care must be taken when fixing the initial conditions. In situations where the cosmological parameters are chosen to exactly coincide with a LCDM cosmology at $z=0$, the high redshift behaviour deviated greatly from LCDM and in fact corresponded to a model dominated by dark-radiation (the effective equation of state was equal to $1/3$). It was found that viable cosmological evolutions were found if one rather fixed the initial conditions to coincide with LCDM at high redshift and evolve the model towards the present time ($z=0$). Care also needs to be taken to avoid sudden curvature singularities, where the cosmographic parameters diverge at finite redshift \cite{Frolov}. It will be possible to explore in detail where in parameter space these pathologies occur using our new Dynamical Systems approach. All these issues will be addressed in a future paper. 

Finally it is worth mentioning that the framework could also be used to compare different inflationary models that produce a scale-invariant power spectrum. Moreover, it is also possible to use the same approach to perform dynamical analysis of other modified gravity theories, \emph{e.g.,} scalar-tensor theories.

\begin{acknowledgments}
PKSD thanks the First Rand Bank for financial support. KM thanks the University of Cape Town for financial support. This work is based on the research supported in part by the National Research Foundation
of South Africa (Grant Numbers: 123055).
\end{acknowledgments}

\appendix

\bibliographystyle{unsrt}
\bibliography{apssamp}

\end{document}